\documentclass{optica-article}
\journal{opticajournal} 
\articletype{Research Article}

\usepackage{graphicx}
\usepackage{placeins}
\usepackage{dcolumn}
\usepackage{bm}
\usepackage{xcolor}
\usepackage{hyperref}
\usepackage{nicefrac}
\usepackage{listings}

\definecolor{codegreen}{rgb}{0,0.6,0}
\definecolor{codegray}{rgb}{0.5,0.5,0.5}
\definecolor{codepurple}{rgb}{0.58,0,0.82}
\definecolor{backcolour}{rgb}{0.95,0.95,0.92}

\lstdefinestyle{mystyle}{
    backgroundcolor=\color{backcolour},   
    commentstyle=\color{codegreen},
    keywordstyle=\color{magenta},
    numberstyle=\tiny\color{codegray},
    stringstyle=\color{codepurple},
    basicstyle=\ttfamily\footnotesize,
    breakatwhitespace=false,         
    breaklines=true,                 
    captionpos=b,                    
    keepspaces=true,                 
    numbers=left,                    
    numbersep=5pt,                  
    showspaces=false,                
    showstringspaces=false,
    showtabs=false,                  
    tabsize=2
}

\usepackage{xspace}

\newcommand{\fmodesep}{2\nu_m}
\newcommand{\errorterm}{\mathcal{O}}
\newcommand{\qlaser}{q_\text{In}}
\newcommand{\qcav}{q_\text{Arm}}
\newcommand{\qrc}{q_\text{RC}}

\newcommand{\Finesse}{\textsc{Finesse3}\xspace}

\newcommand{\HG}[1]{HG$_{#1}$}
\newcommand{\LG}[1]{LG$_{#1}$}
\newcommand{\Linput}{L_\text{IO}}
\newcommand{\Lpd}{L_\text{PD}}
\newcommand{\gouyPD}{\Psi_{\text{PD}}}

\newcommand{\dblinefill}{%
\leavevmode \cleaders \hbox to.5em{\hss=\hss}\hfill\kern0pt
}

\newcommand{\uwa}{Department of Physics, University of Western Australia, Crawley, 6009, Australia}
\newcommand{\ozgrav}{OzGrav, Australian Research Council Centre of Excellence for Gravitational Wave Discovery, Australia}
\newcommand{\ua}{School of Physical Sciences, University of Adelaide, Adelaide, 5005, Australia}

\hypersetup{
    colorlinks,
    linkcolor={red!50!black},
    citecolor={blue!50!black},
    urlcolor={blue!80!black}
}

\begin{document}

\title{Single and coupled cavity mode sensing schemes using a diagnostic field}

\author{Aaron~W. Goodwin-Jones,\authormark{1,2,*}
Haochen Zhu,\authormark{1,2}
Carl Blair,\authormark{1,2}
Daniel D. Brown,\authormark{1,3}
Joris~van~Heijningen,\authormark{1,2,4},
Li Ju,\authormark{1,2}
Chunnong Zhao\authormark{1,2}}
\address{\authormark{1}\ozgrav\\ \authormark{2}\uwa\\ \authormark{3}\ua\\
\authormark{4}Centre for Cosmology, Particle Physics and Phenomenology, Universit\'e catholique de Louvain,
Louvain-La-Neuve, B-1348, Belgium}
\email{\authormark{*}aaron.jones@ligo.org} 

\date{\today}

\begin{abstract*}
Precise optical mode matching is of critical importance in experiments using squeezed-vacuum states. Automatic spatial-mode matching schemes have the potential to reduce losses and improve loss stability. However, in quantum-enhanced coupled-cavity experiments, such as gravitational-wave detectors, one must also ensure that the sub-cavities are also mode matched. We propose a new mode sensing scheme, which works for simple and coupled cavities. The scheme requires no moving parts, nor tuning of Gouy phases. Instead a diagnostic field tuned to the HG20/LG10 mode frequency is used. The error signals are derived to be proportional to the difference in waist position, and difference in Rayleigh ranges, between the sub-cavity eigenmodes. The two error signals are separable by 90\,degrees of demodulation phase. We demonstrate reasonable error signals for a simplified Einstein Telescope optical design. This work will facilitate routine use of extremely high levels of squeezing in current and future gravitational-wave detectors.
\end{abstract*}



%
\section{Introduction}
Precise optical mode matching is of critical importance to many experiments, especially those that are sensitive to optical loss. Examples include Free Space Optical Communications~\cite{Stotts21,Facebook_Air_Ground_Link}, CV-QKD (e.g. \cite{Pirandola2020}) and Quantum Teleportation (e.g. \cite{Simon2017}). Optical setups using higher-order modes (e.g. \cite{Zeng18, Jiao20, Mestre10}) are particularly sensitive to optical loss \cite{Jones20a}.

One particularly interesting problem is posed by ground-based Interferometric gravitational-wave detectors (GWDs) \cite{geo600_6db,Tse2019,Acernese2019}. In these GWDs high-power coherent and squeeze states are coupled between several resonators with low loss. The problem is further complicated by high optical power effects degrading the mode matching \cite{vinet09,Rocchi11,Brooks16}. 

Several techniques have been proposed to directly detect waist position and waist radius mismatches, such as bulls-eye photo-detection \cite{mueller2000}, mode converters \cite{fabian_qpd} and radio frequency (RF) beam shape modulation \cite{ciobanu2020}. In addition, several cameras have been proposed to decompose the coupled cavity mode content in an arbitrary basis \cite{cao2019, Schiworski:21, Ralph17,KSWD07,Takeno11,Bogan15,Jones20,Jones21}. However, to the authors knowledge no sensor directly interrogates the \textit{coupled-cavity} mismatch.

In this work, we extend an early alignment sensing proposal \cite{anderson84} with recent experimental ideas \cite{ciobanu2020} to develop a new mode sensing scheme. In contrast to \cite{mueller2000,fabian_qpd,ciobanu2020}, the scheme requires only a small photodiode and a phase-modulator. This scheme is summarized in \S\,\ref{sec:fp}.

Furthermore, of particular interest to the GWD community, this scheme directly interrogates the mismatch between the eigenmodes in a \textit{coupled-cavity}. Herein we refer to this as the coupled cavity mismatch. We discuss the application of this scheme to a coupled cavity in \S\,\ref{sec:cc}. As a toy example, we consider a recycling cavity and arm cavity based on the proposed Einstein Telescope core interferometer in \cite{Rowlinson21}. 

In \S\,\ref{sec:four_deg} we discuss simultaneous matching of the input beam, recycling cavity and arm cavity. We then briefly discuss advantages and implementation details of this scheme in the context of a GWD. Lastly \S\,\ref{sec:summary} contains a summary and outlook. A table of frequently used symbols is contained in App.~\ref{app:notation}.

\section{A Simple Fabry-Perot cavity}
\label{sec:fp}
The scheme we propose uses two new elements. The first is a field at the \HG{20} mode frequency, in an an-astigmatic cavity this is equal to the \LG{10} (LGpl) \cite{Beijersbergen93}. The second is an optical device to mix these fields and extract the error signal.
\FloatBarrier
\begin{figure}
    \centering
    \includegraphics[width=0.7\linewidth]{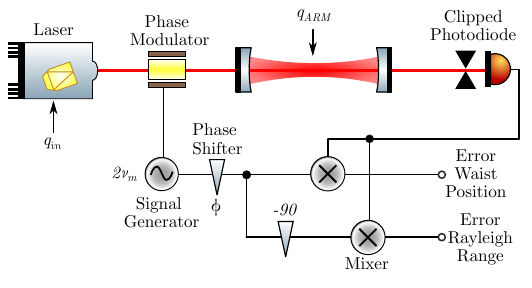}
    \caption{Illustration of mode sensing scheme with a simple Fabry-Perot cavity}
    \label{fig:fp_schematic}
\end{figure}
\subsection{Injection}
The additional field may be excited with a phase modulator, which produces two frequency sidebands. We are only interested in the lower (negative) sideband. For modulation index, $g$, driven at the second order mode frequency, $\fmodesep$\,Hz, the amplitude of the lower sideband, at the cavity input mirror, will be \cite{Bond2017}
\begin{align}
    a_i = \frac{i\sqrt{P}g}{2}\exp(-2i\pi |\fmodesep| \Linput).
    \label{eq:ai}
\end{align}
$\Linput$ denotes the distance between the phase modulator and the cavity we want to mode match to. We assume the light has complex beam parameter $\qlaser$ and use the Hermite-Gaussian modes to describe the beam.

The beam parameter of the cavity fundamental mode is defined as $q_{arm}$ and we consider the case where there is a small discrepancy between the input beam parameter $q_{in}$ and $q_{arm}$. To calculate the magnitude of scatter from the input beam to the modes of the circulating beam $k_{00nm}$ we assume a mismatching in one transverse component x. The scatter coefficient from the fundamental mode to the second order transverse mode $k_{0020}$ is (Table 1, \cite{ciobanu2020}),
\begin{align}
k_{0020} &\approx \frac{\sqrt{2}(i\Delta_z - \Delta_{zR})}{4\overline{z_R}} + \errorterm(\Delta_z^2,  \Delta_{zR}^2)\label{eq:k0020_approx}.
\end{align}
$\Delta z$ and $\Delta z_R$ denote the differences in waist position and Raleigh range between the two eigenmodes. The average rayleigh range is $\overline{z_R} = (z_{R,1} + z_{R,2})/2$. The amplitude of the sideband, in the \HG{20} mode, on the cavity side of the input mirror is then
\begin{align}
    b_0 = \beta_0 k_{0020}\text{ for } \beta_0 \equiv \frac{ia_it_0\exp(-2i\Psi(\qcav; z_0))}{1-r_0r_1\exp(i\phi_\text{r.t.})}. \label{eq:a0_simple_cav}
\end{align}
$\beta_0$ is a convenient phasor. The power reflectivity of the input and output mirrors are $r_0^2$ and $r_1^2$ and the power transmission of the mirrors are $t_0^2$ and $t_1^2$. If the modulator is driven at the second order mode frequency, then the positive Gouy phase accumulation exactly cancels the negative sideband phase accrual. Thus, if the cavity is held on resonance for the carrier light, then the round trip phase accrued is $\phi_\text{rt}=0$. $\Psi(\qcav; z_0)$ is the instantaneous Gouy phase of $\qcav$, evaluated at the input mirror and is sometimes expressed as part of Eq.~\ref{eq:k0020_approx}. Eq.~\ref{eq:a0_simple_cav} states that the resonant field in the cavity is linear with waist position and waist radius mismatch. Furthermore, these fields are separated in phase by 90 degrees. 
\subsection{Sensing}
\label{sec:sensing}
The field at the clipped photodetector is then,
\begin{align}
b_t = \beta_t\beta_0 k_{0020}\text{ for }\beta_t\equiv t_2\exp\left(i \left[ 2\gouyPD - \frac{2\pi(\fmodesep)}{c}\Lpd \right]\right),
\label{eq:bt}
\end{align}
where $\Lpd$ is the distance between the input coupler and photo-diode, while $\gouyPD$ is the Gouy phase accumulated over this path. 

Both the Hermite-Gaussian and Laguerre-Gaussian modal basis are orthonormal, thus, the mode $m,n+2$ mixed with the mode $m,n$ will not produce a time varying signal on an ideal infinite aperture photodetector. However, a finite aperture will break the orthogonality of the modes, thus producing a beat note. Formally, we calculate this in two steps. First, by computing the scatter from the transmitted, carrier \HG{00}, $a_t$, into carrier (\HG{00}) after the aperture. This  models the \HG{00} power lost on the aperture surface. For aperture of radius $A$ (normalized by the beam radius), the result is
\begin{equation}
    k^A_{0,0,0,0} = (1-\exp(-2A^2))\exp(i\gamma_{0,0,0,0}),
    \label{eq:kA0000}
\end{equation}
as derived in App.\,\ref{app:overlap}. The factor $\gamma_{n,m,0,0}$ is related to the instantaneous Gouy phase, $\Psi$, as
\begin{align}\label{eq:gamma_nm00}
    &\exp(i\gamma_{n,m,0,0}) \equiv\Big(\sqrt{\exp(i(2n+1)\Psi)}\Big)^*\sqrt{\exp(i\Psi)} 
    \Big(\sqrt{\exp(i(2m+1)\Psi)}\Big)^*\sqrt{\exp(i\Psi)}.
\end{align}
At the waist, or in the far field, $\gamma_{n,0,0,0}=0$ for $n\in \{0,2\}$. Second, we compute the scatter from \HG{00}, at the lower sideband frequency, before the aperture into \HG{00}, at the lower sideband frequency, after the aperture (also derived in App.~\ref{app:overlap}), 
\begin{align}
    k^A_{2,0,0,0} &= -\exp(i\gamma_{2,0,0,0})\sqrt{2}A^2\exp(-2A^2).
    \label{eq:kA2000}
\end{align}
Considering these functions, an aperture size $A\sim\nicefrac{3}{4}$ is reasonable. As discussed in App.\,\ref{app:overlap}, the optimal choice depends on whether you are shot noise, or cross-coupling limited. In this work we use $A=0.73$, $\therefore$ $k^A_{2,0,0,0}k^A_{0,0,0,0}$ $\approx -0.17\exp(i\sum_{n\in\{0,2\}}\gamma_{n,0,0,0})$. 

The final step is to compute the beat strength. For $k_{0020}\ll 1$, the transmitted power, $P_T$, is dominated by the resonant HG00 mode therefore $\sqrt{P_t}\approx a_t$. We specify a complex beat-coefficient, 
\begin{align}
\label{eq:pbeat}
P_\text{Beat} &\approx -0.17 \times e^{i\sum_{n\in\{0,2\}}\gamma_{n,0,0,0}} \sqrt{P_T}\beta_t\beta_0\frac{i\Delta_z - \Delta_{zR}}{\overline{z_R}}
\end{align}
valid for $A\approx 0.73$. Since we have tracked the phases of all fields relative to the \HG{00} mode, the phase of $a_t$ is not relevant and the phase of $P_\text{Beat}$ does not depend on the microscopic positioning of the detector. The appropriate demodulation phase and gain can be found by evaluating Eq.~\ref{eq:pbeat} for $0<\Delta z\ll z_R$. In any real implementation, $P_\text{Beat}$ would need to be multiplied by the photodiode responsivity and transimpedance gain to obtain some measurable voltage. 

\subsection{Example}
Consider a 750\,mm-long, symmetric impedance-matched cavity with 500\,mm radii of curvature mirrors. The modulator is 100\,mm from the cavity and a clipped photo-diode is placed a further 100\,mm from the cavity transmission, with aperture 0.727$w(z)$. The optical layout is shown in Fig. \ref{fig:fp_schematic}.
\begin{figure}
    \centering
    \includegraphics[width=0.7\linewidth]{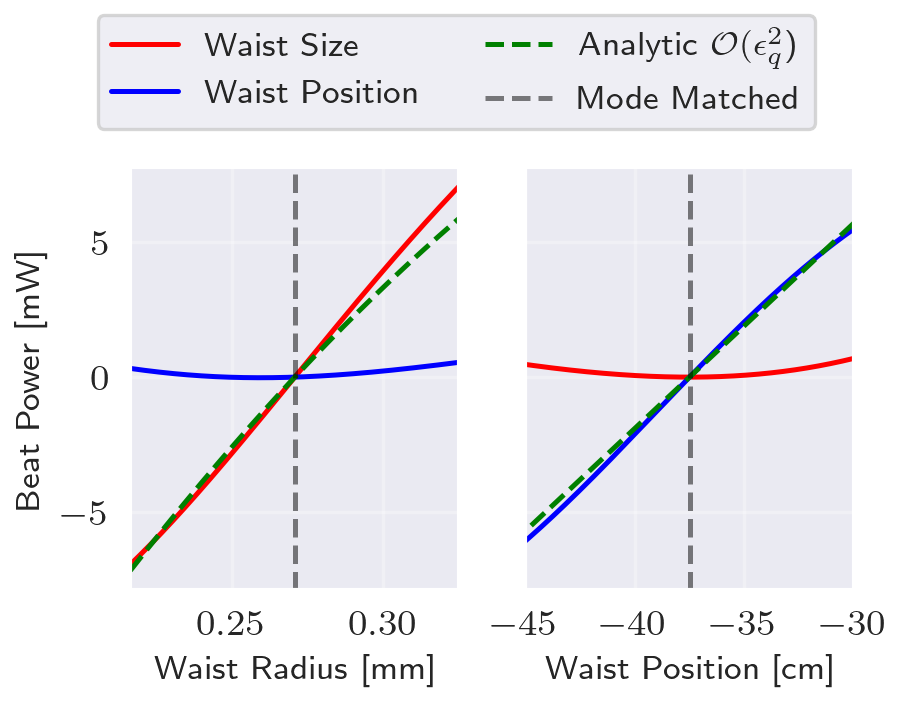}
    \caption{Numerically modelled and analytic error signals for the setup shown in Fig. \ref{fig:fp_schematic}. The analytic expression was obtained by evaluating Eq.~\ref{eq:pbeat}. The non-linearity in the analytic expression arises due to the implicit non-linearity in the denominator of Eq.~\ref{eq:k0020_approx}. At high mismatch two effects cause cross-coupling and non-linearity in the simulated data. Firstly, shifting Gouy phases change the appropriate demodulation phase, causing cross-coupling between the waist position and radius channels. Secondly, the higher order terms in Eq.~\ref{eq:k0020_approx} become significant. When the system is close to mode-matched, error signals are linear, in agreement and without cross-coupling.}
    \label{fig:fp_results}
\end{figure}

The system was modelled using the \Finesse frequency domain numerical library \cite{Finesse3} and compared against the analytics derived earlier. See \S\,\ref{app:general_comments_moddelling} for a brief discussion on modelling details. The result is shown in Fig.~\ref{fig:fp_results}. The laser input power is 1\,W and the modulation index 0.1. Near to the operating point, the error signal is linear. Far from the operating point, the approximations derived earlier break down.

Crucially, and uniquely the error signal has a null when the system is correctly mode matched. Therefore the scheme can be used to verify the operating point and verify the target of other mode matching sensors such as the Hartmann sensor at LIGO (\cite{Brooks16} \& therein).

\section{A Coupled Cavity}
\label{sec:cc}
\begin{figure}
    \centering
    \includegraphics[width=0.7\linewidth]{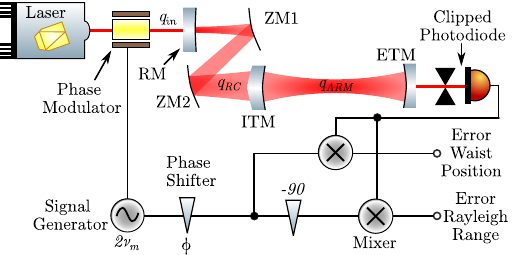}
    \caption{Illustration of mode sensing scheme with a coupled cavity. An incoming beam, $\qlaser$ passes through an EOM which adds a pair of frequency sidebands at $\fmodesep$\,Hz. The beam passes through a Recycling Mirror (RM) and is then expanded by two focusing optics (ZM1 \& ZM2). Since $\fmodesep$ is less than the FWHM of the recycling cavity, it is amplified by the recycling cavity and converted into $\qrc$. Mismatches}
    \label{fig:cc_schematic}
\end{figure}
\begin{figure}
    \centering
    \includegraphics[width=0.6\linewidth]{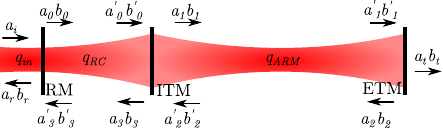}
    \caption{Simplified analytic model describing Fig~\ref{fig:cc_schematic}. Focusing optics are ommited and a gaussian eigenmode is assumed to exist in both the arm and recycling cavities. $a$ fields are in the HG00 mode shape, $b_{\{\}}$ fields are in the HG20 mode shape. $x_{i,r,0,3}$ are in the $\qrc$ basis and $x_{1,2,t}$ are in the $\qcav$ basis. For details on the phase acrued by each mode, please see text.}
    \label{fig:cc_fields}
\end{figure}
We now adapt the scheme to directly sense the mismatch between the eigenmodes of a coupled cavity. Adopting gravitational wave nomenclature we will refer to the cavity nearest the laser Recycling Cavity (RC) and the cavity furthest from the laser as the arm cavity (ARM). We assume that the full width at half maximum of the RC is much greater than the FSR of the arm. This is true when the ARM is significantly longer and higher finesse than the recycling cavity. 

As in \S\,\ref{sec:fp}, we consider only the field at $\fmodesep$ and assume some frequency/length stabilization tunes both cavities to maximize power in the arm. We define the leftmost mirror in Fig.~\ref{fig:cc_fields} to have amplitude transmission $t_0$ and reflection $r_0$, the middle mirror to have $t_1,r_1$ and the rightmost mirror to have $t_2,r_2$. We define the input beam, $\qlaser$ to be well matched to the eigenmode in the RC, $\qrc$. We assume a small mismatch exists between $\qrc$ and $\qcav$. The round trip phase accumulated, in the RC, by the \HG{00} is $\phi_\text{RC}^a$ and the Arm is $\phi_\text{ARM}^a$. The \HG{20} accumulates $\phi_{\{\text{RC},\text{ARM}\}}^b$ and \cite{bayer-helms},
\begin{align}
   k_{n,m,n',m'} &= \int\int u_{n,m}^*(\qrc)u_{n',m'}(\qcav).
   \label{eq:knmnm_general}
\end{align}
From Table 1 in \cite{ciobanu2020}, we find,
\begin{align}
k_{2020} &\approx 1 - \frac{5i\Delta_z}{4\overline{z_R}} + \errorterm(\Delta_z^2,\Delta_{zR}^2)\approx 1 \text{ for } \Delta_z\ll z_R \label{eq:k2020_approx}\\
k_{0000} &\approx 1 - \frac{i\Delta_z}{4\overline{z_R}} + \errorterm(\Delta_z^2,\Delta_{zR}^2)\approx 1 \text{ for } \Delta_z\ll z_R \label{eq:k0000_approx}\\
k_{2000} &\approx k_{0020}^* \approx \frac{\sqrt{2}(-i\Delta_z - \Delta_{zR})}{4\overline{z_R}} + \errorterm(\Delta_z^2,  \Delta_{zR}^2)\label{eq:k2000_approx}
\end{align}
As shown in Fig.~\ref{fig:cc_fields}, there are four non-trivial fields at the lower sideband frequency. $a_3$ and $a_1$ refer to the \HG{00} mode in the recycling cavity and arm respectively. $b_3$ and $b_1$ refer to the \HG{20} mode in the respective cavities. Neglecting mirror surface roughness, the coupling between the resonators is given by
\begin{align}
a_1 &= a'_2 r_1 + i t_1(k_{0000} a'_0 + k_{2000}b'_0) \label{eq:a1}\\
a_3 &= a'_0 r_1 + i t_1(k_{0000} a'_2 + k_{2000}b'_2) \label{eq:a3}\\
b_1 &= b'_2 r_1 + i t_1(k_{2020} b'_0 + k_{0020}a'_0) \label{eq:b1}.\\
b_3 &= b'_0 r_1 + i t_1(k_{2020} b'_2 + k_{0020}a'_2) \label{eq:b3}
\end{align}
The numerical model, \Finesse, is able to solve these equations fully. However, they are cumbersome to solve analytically and so we first solve for the \HG{00} mode neglecting coupling from the \HG{20} back into the \HG{00}.

For small mismatch, Eqs.~\ref{eq:k0000_approx} and \ref{eq:k2000_approx} state that, $|k_{2000}| \ll |k_{0000}| $. Additionally, we assume that the \HG{20} field does not resonate in the RC since the RC mode separation frequency $\gg$ than the RC FWHM. Therefore $|k_{2000}b'_0| \ll |k_{0000} a'_0|$ and Eq.~\ref{eq:a1} may approximated as,
\begin{align}
a_1 &\approx a'_2 r_1 + i t_1k_{0000} a'_0.  \label{eq:a1_approx}
\end{align}
Next we assume that the $|k_{0000} a'_2| \gg |k_{2000}b'_2|$. As $k_{2000}\rightarrow 0$ this will be true. 
Qualitatively, we are assuming that the coupling between the off resonance \HG{00} mode in arm cavity is larger in amplitude than the coupling into and back out of the resonant \HG{20} mode. This is not generally true, but becomes true as the coupling between the modes tends to zero, $k_{0020}\rightarrow0$. For further discussion see App.~\ref{app:discussion_a0}. 
When this approximation fails, the appropriate gain and demodulation phase will not be predicted by the analytic expressions. When true, Eq.~\ref{eq:a3} may be rewritten as,
\begin{align}
    a_3 &\approx \label{eq:a3_approx} a'_0 r_1 + i t_1k_{0000} a'_2.
\end{align}
\begin{figure}
    \centering
    \includegraphics[width=0.5\linewidth]{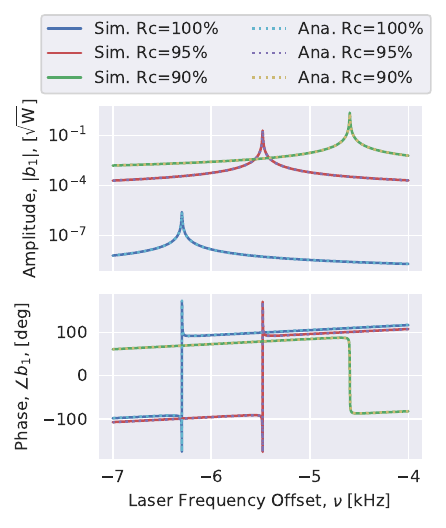} 
    \caption{The phase and amplitude of the HG20 mode, evaluated at the surface of the ITM, $b_1$. Solid lines show a simulation and dashed lines show Eq.~\ref{eq:b1sol}. The result is computed for a 100\,\%, 95\,\% and 90\,\% of the nominal ITM RoC. These correspond to mismatches of 0, 360\,ppm and 5.7\,\% respectively.}
    \label{fig:b0}
\end{figure}
As shown in App.~\ref{app:cc_deriv}, the trivial fields can then be solved to reveal $a_0$ in terms of $a_i$,
\begin{align}
    a_{0}\approx\frac{i a_{i} t_{0} \left(- r_{1} r_{2} e^{i \phi_\text{ARM}^a} + 1.0\right)}{k_{0000}^{2} r_{0} r_{2} t_{1}^{2} e^{i \left(\phi_\text{RC}^a + \phi_\text{ARM}^a\right)} + r_{0} r_{1}^{2} r_{2} e^{i \left(\phi_\text{RC}^a + \phi_\text{ARM}^a\right)} - r_{0} r_{1} e^{i \phi_\text{RC}^a} - r_{1} r_{2} e^{i \phi_\text{ARM}^a} + 1.0}.
    \label{eq:a0}
\end{align}
Finally we assume $k_{2020} b'_2 \gg k_{0020}a'_2$, which is also true as $k_{0020} \rightarrow 0$. Once again, when the approximation breaks down, the gain and demodulation phase will change. When the approximation is true, Eq.~\ref{eq:b3} may be rewritten as,
\begin{align}
b_3 &= \approx b'_0 r_1 + i t_1 k_{2020} b'_2 \label{eq:be_approx}
\end{align}
One may repeat this process to determine the resonating HG20 field in the Arm, as a function of the amplitude of the HG00 field in the RC ($a_0$). As shown explicitly in Appendix~\ref{app:bcc_deriv} it is,
\begin{align}
    b_{1}=- \frac{i a_{0} k_{0020} t_{1} \left(r_{0} r_{1} e^{i \phi_\text{RC}^b} - 1\right) e^{\frac{i \phi_\text{RC}^a}{2}}}{k_{2020}^{2} r_{0} r_{2} t_{1}^{2} e^{i \left(\phi_\text{RC}^b + \phi_\text{ARM}^b\right)} + r_{0} r_{1}^{2} r_{2} e^{i \left(\phi_\text{RC}^b + \phi_\text{ARM}^b\right)} - r_{0} r_{1} e^{i \phi_\text{RC}^b} - r_{1} r_{2} e^{i \phi_\text{ARM}^b} + 1}
    \label{eq:b1sol}
\end{align}
\subsection{Validation of the Coupled Cavity Equations}
To validate the approximations made in the preceding paragraph, we use \Finesse to construct a numerical model of a coupled cavity. \Finesse fully models Eq.~\ref{eq:a3}-\ref{eq:b1}. The optical system we choose is based on the proposed Einstein Telescope with beam expanding telescopes in the arms \cite{Rowlinson21}. For more details see Appendices~\ref{app:general_comments_moddelling} \& \ref{app:cc_design}. The incoming beam is a perfect HG00, mode matched to the first cavity with 1\,W of power.

We compute the field $b_1$ as a function of laser frequency offset, $\nu$ for several ITM Radii of Curvature (RoC) and compare this against simulation. The simulation is such that $\nu=0$\,Hz and nominal RoC results in maximum arm-cavity power build up, 12\,kW. Thus 22\,W of TEM00 in the RC.

As in \S\,\ref{sec:fp}, the phase accrued by a sideband of frequency $\nu$ is $\phi^a_i = -\sum_j 2\pi n_{ji} \nu L_{ji}/ c$, where $n_{ji}$ is the refractive index of material $j$ in cavity $i$ and $L_{ji}$ is the total round-trip path-length of that material in cavity $i$. 
The phase accrued by the HG20 is $\phi_i^b = \phi_i^a + 2\Psi_i$ for round-trip Gouy phase $\Psi_i$.

Fig.~\ref{fig:b0} shows the approximate analytical and full numerical solutions for the nominal design and two designs with shorter ITM RoC. A mismatch like this may occur due to thermal effects in the mirror or due to a design mismatch. One can see excellent agreement between the approximate and numerical solutions. This suggests that approximations made earlier are indeed valid. Discussion on the limitations of these approximations can be found in Appendix~\ref{app:discussion_a0}.
\subsection{A Coupled Cavity Mismatch Sensing Scheme}
\label{sec:cc_app}
\begin{figure}
    \centering
    \includegraphics[width=0.9\linewidth]{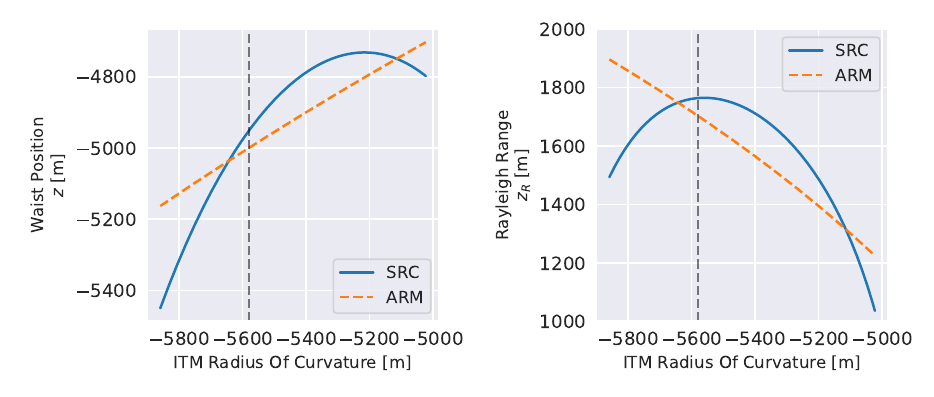} 
    \caption{The distance to the waist, $z$, and Rayleigh range, $z_R$, of the spatial eigenmodes in the RC and Arm. Both eigenmodes have been propagated to a point just in front of the ITM HR surface, enabling a like-for-like comparison. See \S\,\ref{sec:cc_example} for further details.}
    \label{fig:zr_z}
\end{figure}
As in \S \ref{sec:fp}, the sideband is set to the second order mode frequency, therefore $\phi_\text{ARM}^b=0$. We simplify Eq.~\ref{eq:b1sol} and define,
\begin{align}
\beta_{cc} = \frac{b_{1}}{k_{0020}} \approx\label{eq:b1_simp}
\frac{a_{0} t_{1} \left(r_{0} r_{1} e^{i \phi_\text{RC}^b} - 1\right) e^{i(\frac{ \phi_\text{RC}^a}{2}-2\Psi_2)}}{ \left(r_{0} r_{1}^{2} r_{2} e^{i \phi_\text{RC}^b} - r_{0} r_{1} e^{i \phi_\text{RC}^b} + r_{0} r_{2} t_{1}^{2} e^{i \phi_\text{RC}^b} - r_{1} r_{2} + 1\right)}.
\end{align}
Following the arguments in \S \ref{sec:sensing}, the error signal is,
\begin{align}
\label{eq:pbeat_cc}
P_\text{Beat} &\approx -0.17 \times e^{i\sum_{n\in\{0,2\}}\gamma_{n,0,0,0}} \sqrt{P_T}\beta_t\beta_{cc}\frac{i\Delta_z - \Delta_{zR}}{\overline{z_R}}
\end{align}
where $\beta_t$ (defined in Eq.~\ref{eq:bt}) is updated to use the distance and accumulate Gouy phase from the ITM HR surface to the output photo-detector. The above equation clearly shows that we expect an error signal which is linearly proportional to waist position and Rayleigh range, with both degrees seperated by 90 degrees in demodulation phase. However, recall that we have assumed $k_{0020} \rightarrow 0$. When $k_{0020}$ grows, we expect this demodulation phase and amplitude to change as these approximations break down. 
\subsection{Example}
\label{sec:cc_example}
\begin{figure}
    \centering
    \includegraphics[width=.8\linewidth]{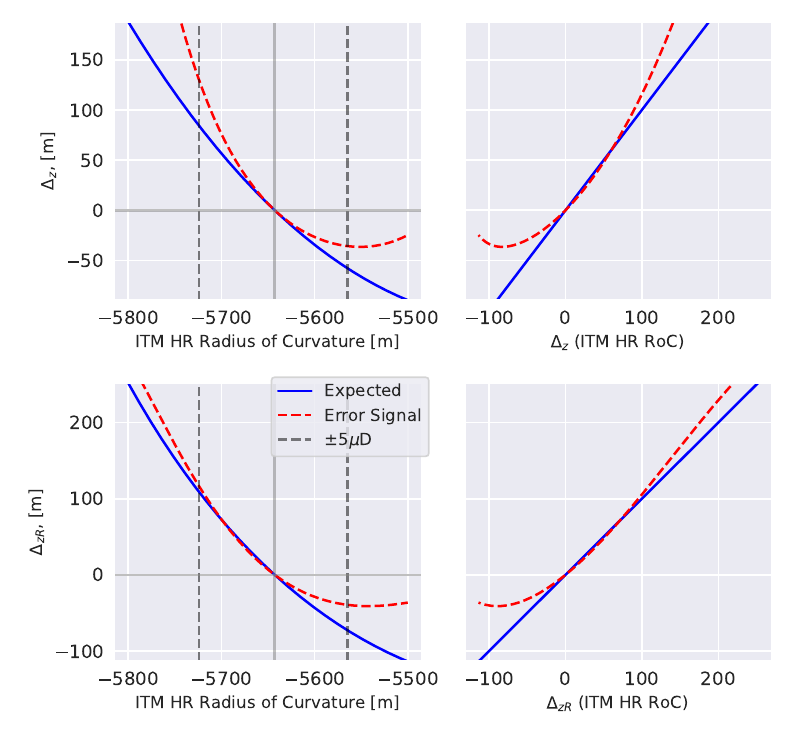} 
    \caption{The left hand column shows error signals as a function of ITM RoC. \textit{Expected} shows the difference between cavity eigenmode waist position and Rayleigh Range, computed from ABCD matrices. \textit{Error Signal} shows the output of a simulated photo-diode, demodulated at $\fmodesep$. The demodulation phase and calibration were obtained from Eq.~\ref{eq:pbeat_cc}. The right hand column is analogous to Fig.~\ref{fig:fp_results} and shows the error signal against the actual difference in waist position and Rayleigh Range, obtained from ABCD matrices.}
    \label{fig:cc_calib}
\end{figure}
We now apply the mode sensing scheme to the realistic cavity shown in Fig.~\ref{fig:cc_schematic}. This is the same optical configuration as was used for \S \ref{sec:cc_app}, where we choose parameters based loosely on the proposed \cite{Rowlinson21} power recycling cavity and arm of the Einstein Telescope Low Frequency interferometer. 
As discussed earlier, we assume the FWHM of the RC is greater than the FSR of the arm. Therefore, we require a beam expansion telescope in the arm \cite{Rowlinson21}. In our case, it is provided by the curved mirrors ZM1 and ZM2. In our example, $\fmodesep \approx 6$\,kHz and the RC FWHM is 7\,kHz.

One source of mismatch in GWDs is thermally induced RoC changes in the coupling mirror \cite{vinet09,Rocchi11,Brooks16}. Fig~\ref{fig:zr_z} shows the RC and Arm cavity eigenmode waist position and Rayleigh range. This is shown as a function of ITM RoC, over a wide range. We use the cavity round trip ABCD matrix to compute the eigenmode. In the case of the Arm, the change is reasonably linear, however the change in the RC has a turning point at around ITM RoC = $-5400$\,m. Around this turning point, 
the round trip Gouy phase of the RC passes through 90 degrees, causing an anti-resonance in the RC for the HG20 mode.

The left hand column in Fig.~\ref{fig:cc_calib} shows $\Delta_z$ and $\Delta_{zR}$ as a function of ITM radius of curvature. The \textit{expected} trace is from the ABCD matrices. We see that a change in the ITM RoC non-linearly changes the mismatch between the cavities, as also observed in Fig.~\ref{fig:zr_z}. The \textit{error signal} trace, shows the output of a simulated clipped photodiode. This output has been calibrated meters by using Eq~\ref{eq:pbeat_cc} to determine the correct demodulation phase and gain.

In this case, the gain was $1.8\times 10^{-9}$\,W/m with a complex phase of $133$ degrees. This sounds small, however, ET-LF has only $100$\,mW of power on transmission. Furthermore, the gain could be increased with appropriate low-noise transimpedance amplification and filtering. Thus, in the upper plot, \textit{Error Signal} shows the output of this photodiode, demodulated with phase $133$ degrees and divided by $1.97\times 10^{-9}$. For the lower plot, one adds $90$ degrees to the demodulation phase to select the Rayleigh Range.

Within $\pm 5$\,\textmu D one can see monotonic agreement the expected trace and the \textit{error signal}. This corresponds to a 80\,m RoC tuning or 200\,km \textit{Thermal Lens}. As we move further from the operating point, several effects become significant. Firstly, the Gouy phase in the cavities will change. For example, in our system the round trip Gouy phase in the Arm goes from 278\,deg. at -5800\,m RoC to 304\,deg at -5000\,m RoC. The change in the RC is more significant going from 128\,deg to 36\,deg over the same range. Secondly, the approximations we made in \S~\ref{sec:cc} begin to breakdown. Finnaly, around ITMX RoC = $-5540$ the HG20 anti-resonates in the PRC causing a change in the sign of of the $b_1$ and hence the demodulation phase. Crucially, the error signal has a null when the system is correctly mode matched. Therefore the scheme can be deployed as a witness to the true sub-cavity mismatch, which is a useful when diagnosic. 

The right hand column, uses the \textit{expected} trace as an $x$ axis for the plot. In this way, the useful range is more clearly highlighted. 

If one has a well characterised cavity, with extensive wavefront sensing and real time diagnostic tools (such as at the LIGO observatories), it may be possible to infer the RoC. The Gouy phases can be computed in real-time and used to produce a more accurate demodulation phase and gain. This was trialled, however, the dynamic range enhancement was only marginal due to the resonances in the RC around $\sim -5000$\,m.
\section{Future work and possible implementation in GWDs}
\label{sec:four_deg}
\begin{figure}
    \centering
    \includegraphics[width=0.7\linewidth]{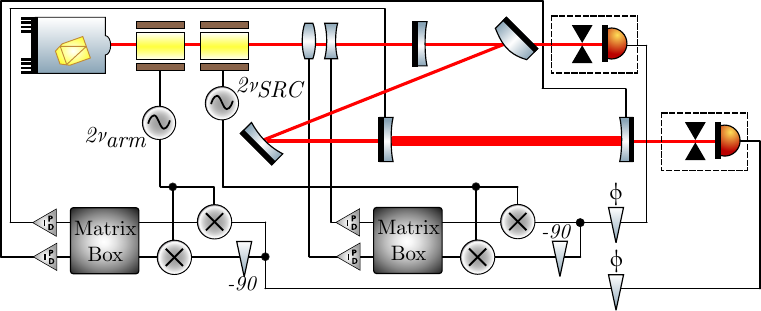}
    \caption{A possible implementation of the \textit{OzGrav} mode sensing scheme, where all four mode-matching degrees of freedom are controlled. One modulator excites a second order mode field at twice the RC mode separation frequency, this is resonant in the RC, but not in the arm and is sensitive to $\qrc = \qlaser$. The second modulator excites an additional field at twice the arm mode-separation frequency. Symbols are as defined in previous figures. See \S\,\ref{sec:four_deg} for discussion.}
    \label{fig:double_cavity}
\end{figure}
In a coupled cavity, there are four canonical mode-matching degrees of freedom. These are the Rayleigh range and waist position difference between the RC and the input mode; and the Rayleigh range and waist position difference between the ARM and the RC.

In \S\,\ref{sec:cc}, we implicitly assumed that $\qrc = \qlaser$ by some automatic control scheme, which is trivially achieved in \Finesse. In a real experiment, one has the choice of several automatic mode-matching schemes \cite{mueller2000,fabian_qpd,ciobanu2020}. Any one of these schemes could be used in conjunction with the presented coupled-cavity scheme to eliminate mismatch along the optical path. Our preferred option, is to apply the presented scheme twice, which is shown schematically in Fig~\ref{fig:double_cavity}.

In this instance, we excite two additional fields. The first field is at twice the RC mode-separation frequency. This causes a HG20 field $\propto \qrc - \qlaser$ to be resonant in the RC. The field may be read from a pick-off port in the RC, in the manner described in \S\,\ref{sec:fp}. Since the field is not resonant in the arm, it is not sensitive $\qcav - \qrc$. The second field is at twice the arm mode separation frequency. This field is is then readout as described in \S\,\ref{sec:cc}. Depending on the precise design of the system, this field may be sensitive to all four mode sensing degrees of freedom. The error signals may be decoupled with a decoupling matrix. A gain hierarchy may also be required for stable control of the four degrees of freedom. 

Within GWDs there are not two, but four coupled cavities. Generally, the most critical to track are mode matches along the squeezing paths \cite{Kwee2014,toyra17,mcculler20}. If the field at $\fmodesep$ was injected in along with the squeezed input, it would resonate in the Signal Recycling Cavity. In LIGO, one option is to use a modified ADF \cite{Dhruva_ADF}. Work is ongoing with LIGO Laboratory to advance the concepts in this work and proof-of-principal measurements have been obtained \cite{Jones22_LLO}.
%
%
%
%
%
%
%
%
\section{Summary and Outlook}
\label{sec:summary}
We present a simple mode sensing scheme which is directly sensitive to the mismatch between two laser beams. Our scheme only requires a phase modulator on the input and a small area photodiode on transmission, in most cavity laser optics experiments these items are already present for longitudinal control. For a Fabry Perot cavity, we derive linear error signals for the mismatch between the incoming beam and the cavity eigenmode. We demonstrate that these error signals are proportional to the difference in waist position and Rayleigh range. We verify our error signals against simulation and discuss how the demodulation phase may be chosen. 

We then extend the concept to directly measure the waist position and Rayleigh range a coupled cavity. We compute the field in the coupled cavity and derive the approximate error signal. We compare our approximate analytic solution to a numerical model and we find good agreement. Furthermore, we then apply our analytic solution to a realistic coupled cavity, proposed for a future GWD. We find that the error signals are linear, provided that the Gouy phase shifts caused by the mismatch are small and the cavity is not significantly mismatched. In the case of our toy system, we find good agreement for up to $5\,$\textmu D of RoC change. 

Finally, we discussed some of the implementation details which would be required to demonstrate this technology at scale.

\begin{backmatter}
\bmsection{Funding}
This research was conducted with the support of the Australian Research Council Centre of Excellence for Gravitational Wave Discovery (CE170100004). D.D.B. is the recipient of an ARC Discovery Early Career Award (DE230101035) funded by the Australian Government.

\bmsection{Acknowledgments}
We wish to thank the developers, documentation team and testers for their work in developing \Finesse.
We wish to thank the LIGO Lab Squeezing team, ET Wavefront Sensing \& Control and LIGO IFOSIM working groups for useful discussions. Specifically, we wish to thank Paul Fulda, Victoria Xu, Valery Frolov, Dhruva Ganapathy \& Alexei Ciobanu for useful discussions.

A.W.G. acknowledges the LIGO Scientific Collaboration Fellows program for supporting their research at LIGO Livingston Observatory.

This document has been assigned LIGO Document Number \href{https://dcc.ligo.org/P2300010}{P2300010}

\textbf{Author Contributions}: {Lead: A.W.G.; Proposer: C.Z.; Simulations: A.W.G, J.vH, D.D.B.; Mathematics: A.W.G, H.Z., Designed the study: A.W.G., C.B., L.J., C.Z.; Writing: A.W.G.; Critical Revisions: C.B., J.vH., C.Z.}

\bmsection{Disclosures}
The authors declare no conflict of interest

\bmsection{Data availability} No data were generated or analyzed in the presented research.
\end{backmatter}
\appendix

%
%
%
\section{Notation}
\label{app:notation}
Table~\ref{tab:symbols} contains a summary of the frequently used notation.
\begin{table}
\centering
\caption{List of frequently used symbols.}
\label{tab:symbols}
\begin{tabular}{ lp{0.6\linewidth} } 
 \hline \hline
 Symbol & Description\\ \hline\hline
 \multicolumn{2}{c}{Latin}\\\hline
 $A$ & Radius of the aperture in the sensing scheme. See \S \ref{sec:sensing}.\\
 $a_{\{j\}}$ & Field in \HG{00} mode at location $j$, see Fig~\ref{fig:cc_fields} for locations.\\
 $b_{\{j\}}$ & Field in \HG{20} mode at location $j$, see Fig~\ref{fig:cc_fields} for locations.\\
 $k_{nmn'm'}$ & Generic scattering coeffeiciant between mode \HG{nm} and \HG{n'm'}, defined Eq.~\ref{eq:knmnm_general}. For $k_{0020}$ see Eq.~\ref{eq:k0020_approx}. For $k_{2020}$ see Eq.~\ref{eq:k2020_approx}.\\
 $k^A_{nmn'm'}$ & Scattering coeffeiciant on aperture of radius $A$. For $k^A_{0000}$ see Eq.~\ref{eq:kA0000}. For $k^A_{2000}$ see Eq.~\ref{eq:kA2000}.\\
 $\Linput$ & Input optical path lenght. See Eq.~\ref{eq:ai}\\
 $\Lpd$ & Output optical path lenght. See Eq.~\ref{eq:bt}.\\
 $P$ & Input Power. \\
  $\qlaser = \qlaser(z)$ & Input complex beam parameter. \\
  $\qrc = \qrc(z)$ & Recycling cavity complex beam parameter. \\
  $\qcav = \qcav(z)$ & Cavity complex beam parameter. \\
  $r_{j}$ & Amplitude reflectivity of mirror, $j$. See Fig~\ref{fig:cc_fields} for definition.\\
  $t_{j}$ & Amplitude transmission of mirror, $j$. See Fig~\ref{fig:cc_fields} for definition.\\
 \hline
 \multicolumn{2}{c}{Greek}\\\hline
 $\beta_{\{\}}$ & Complex phasors, that do not depend to first order on $k_{0020}$. $\beta_0$ is defined in Eq.~\ref{eq:a0_simple_cav}; $\beta_t$ in Eq~\ref{eq:bt} \& $\beta_{cc}$ in Eq.~\ref{eq:b1_simp}.\\
 $\gamma_{n,m,0,0}$ & Phase when \HG{n,m} scatters into \HG{00}, over aperture $A$. Defined in Eq.~\ref{eq:gamma_nm00}.\\
 $\Delta_{z}$ & Difference in the waist position of two eigenmodes.\\
 $\Delta_{zR}$ & Difference in the Rayleigh range of two eigenmodes.\\
 $\fmodesep$ & Second order mode frequency (arm).\\
 $\phi_{\{\text{RC},\text{ARM}\}}^{\{a,b\}}$ & Round trip phase, accumulated by fields $a$ \& $b$ in cavities RC \& ARM. See \S \ref{sec:cc}.\\
 $\Psi=\Psi(z)$ & Instantaneous Gouy phase.\\
 $\gouyPD$ & Gouy phase accrued on output path. See Eq.~\ref{eq:bt}.\\
 \hline \hline
\end{tabular}
\end{table}
\section{Overlap Integral}
\label{app:overlap}
A finite aperture photo-detector will scatter light between modes, thus causing a beat signal between two normally orthogonal modes. Firstly, we consider the scatter from the carrier longitudinal mode in the HG00 mode into the HG00 mode after the aperture, $k^A_{0000}$. The scatter between modes is given by the overlap integral \cite{bayer-helms},
\begin{align}
    k^A_{n,m,0,0} = \int\int_A &u^*_0(x,z) u_n(x,z)\nonumber\\&u^*_0(y,z) u_m(y,z) \mathrm{d}S.
    \label{eq:knm00}
\end{align}
First, we may collect all the real terms of the HG mode distribution function with index $h$ and define,
\begin{align}
    u'_h(x,z) = &\Big(\frac{2}{\pi} \Big)^{\frac{1}{4}} \sqrt{\frac{1}{2^hh!w(z)}} \nonumber\\
&H_h\Big( \frac{\sqrt{2}x}{w(z)}\Big)\exp\Big(-\frac{x^2}{w^2(z)}\Big).
\end{align}
Then, the complete spatial mode distribution function is (equivalently defined in \cite{Bond2017}),
\begin{align}
    u_h(x,z) = &\sqrt{\exp(i(2h+1)\Psi)} \nonumber\\
    &\exp\Big( -\frac{ikx}{2R_C(z)} \Big) u_h'(x,z).
\end{align}
Since $u'_h(x,z)$ is real,
\begin{align}
    u_h^*(x,z)u_k(x,z) = \Big(\sqrt{\exp(i(2h+1)\Psi)}\Big)^*\nonumber &\\ \sqrt{\exp(i(2k+1)\Psi)} u_h'(x,z)u_k'(x,z)&,
\end{align}
which is true for any $h,k$ located any distance from the waist. We then gather the complex terms into a mode mixing phase. Since all points lie on the complex unit circle; the square roots, conjugates and products also lie on the unit circle. 
\begin{align}
    &\exp(i\gamma_{n,m,0,0}) \equiv \Big(\sqrt{\exp(i(2n+1)\Psi)}\Big)^*\sqrt{\exp(i\Psi)} \nonumber\\
    &\Big(\sqrt{\exp(i(2m+1)\Psi)}\Big)^*\sqrt{\exp(i\Psi)}.
    \tag{\ref{eq:gamma_nm00} Repeated}
\end{align}
Substituting into Eq.~\ref{eq:knm00} gives,
\begin{align}
    k^A_{n,m,0,0} =& \exp(i\gamma_{n,m,0,0})\int\int_A \\
    &u_0'(x,z) u_n'(x,z)u_0'(y,z) u_m'(y,z) \mathrm{d}S \nonumber.
    \label{eq:knm00_reduced}
\end{align}
\FloatBarrier
\begin{figure}
    \centering
    \includegraphics[width=0.7\linewidth]{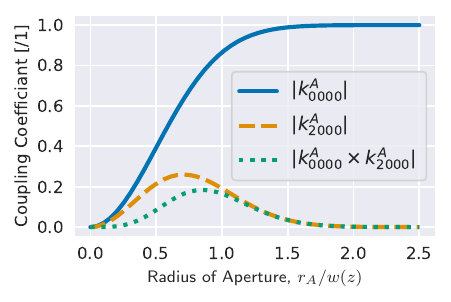}
    \caption{The absolute value of the mode scattering function, $k^A_{n,m,0,0}$, for a mode HGnm, incident on a circular aperture of radius, $r_A$, and beam radius, $w(z)$, scattering into a HG00 mode. As derived exactly in Appendix~\ref{app:overlap}, an aperture with radius $A\sim \nicefrac{3}{4}$ of the beam radius is good for mixing the modes.}
    \label{fig:overlap}
\end{figure}
Now, we compute $k^A_{0,0,0,0}$. Since the aperture is cylindrically symmetric with radius $r=r_A$ polar co-ordinates are more natural. Hence,
\begin{align}
    &k^A_{0,0,0,0} = \exp(i\gamma_{0,0,0,0})\int_0^{r_A}\int_0^{2\pi} \\
    & \exp\left(\frac{-2r^2\left(\cos^2(\theta)+\sin^2(\theta)\right)}{w^2(z)}\right)\frac{2r}{\pi w^2(z)}\mathrm{d}\theta\mathrm{d}r \nonumber.
\end{align}
Solving this integral and substituting $A = r_A/w(z)$, 
\begin{align}
    k^A_{0,0,0,0} = (1-\exp(-2A^2))e^{i\gamma_{0,0,0,0}},
    \tag{\ref{eq:kA0000} Repeated}
\end{align}
which is trivially maximal at $A=\infty$. Eq.~\ref{eq:kA0000} is plotted in Fig.~\ref{fig:overlap}. For the second order mode scatter, we find
\begin{align}
     k^A_{2,0,0,0} &= \exp(i\gamma_{2,0,0,0})\int_0^{r_A}\int_0^{2\pi} \nonumber\\
    &\exp\left(\frac{-2r^2\left(\cos^2(\theta)+\sin^2(\theta)\right)}{w^2(z)}\right)\nonumber \\
    &\left( \frac{8r^2\cos^2(\theta)}{w^2(z)} -2\right)\frac{r}{\sqrt{2}\pi w^2(z)} \mathrm{d}\theta\mathrm{d}r.   
\end{align}
The solution to the integral is,
\begin{align}
    k^A_{2,0,0,0} &= -e^{i\gamma_{2,0,0,0}}\sqrt{2}A^2\exp(-2A^2),
    \tag{\ref{eq:kA2000} Repeated}
\end{align}
which is also plotted in Fig.~\ref{fig:overlap}. It is possible to show that the maximum scattering from HG20 into HG00 occurs when,
\begin{align}
    A=\frac{1}{\sqrt{2}}\: \therefore \: \left|k^A_{2,0,0,0}\right| = \frac{1}{\sqrt{2}\exp(1)}\approx 0.26,
\end{align} by looking for stationary points. However, the stationary point for $k^A_{2,0,0,0}k^A_{0,0,0,0}$ occurs when, 
\begin{align}
  A\approx 0.85 \: \therefore \: \left|k^A_{2,0,0,0}\right| \approx 0.18.  
\end{align}
If the photo-detector is shot-noise limited, the best performance would be obtained by maximising the total beat strength, thus maximising $k^A_{2,0,0,0}k^A_{0,0,0,0}$. However, if the detector is limited by cross-talk from unwanted modulations in the HG00 mode, they can be suppressed by choosing a smaller aperture which maximises $k^A_{2,0,0,0}$ and slightly suppresses $k^A_{0,0,0,0}$.

\section{HG00 Coupled Cavity Derivation}
\label{app:cc_deriv}
We begin by computing the the field $a'_2$ in terms of $a_0$. Using the symbols defined in the main text,
\begin{align}
a'_0 &= a_0\exp (i\phi_\text{RC}^a/2)\\
a_1 &\approx a'_2 r_1 + ia'_0k_{0000}t_1 \tag{\ref{eq:a1_approx} Repeated}\\
a'_1 &= a_1\exp (i\phi_\text{ARM}^a/2)\\
a_2 &= a'_1r_2 \\
a'_2 &= a_2\exp (i\phi_\text{ARM}^a/2).
\end{align}
Solving this system of equations gives:
\begin{align}
a'_{2}\approx- \frac{i a_{0} k_{0000} r_{2} t_{1} e^{i \left(\phi_\text{RC}^a/2 + \phi_\text{ARM}^a\right)}}{r_{1} r_{2} e^{i \phi_\text{ARM}^a} - 1.0}.
\label{eq:a2__approx}
\end{align}
One may then compute,
\begin{align}
a_3 &\approx a'_0 r_1 + it_1k_{0000} a'_2 \tag{\ref{eq:a3_approx} Repeated}\\
a'_3 &= a_3\exp (i\phi_\text{RC}^a/2)\\
a_0 &= a'_3r_0 + it_0a_i.
\end{align}
Solving this system of equations yeilds Eq.~\ref{eq:a0}.
\section{HG20 Coupled Cavity Derivation}
\label{app:bcc_deriv}
Similar to Appendix~\ref{app:cc_deriv}, we begin by computing $b_0$ in terms of $b_1$,
\begin{align}
b'_1 &= b_1\exp (i\phi_\text{ARM}^b/2)\\
b_2 &= b'_1r_2 \\
b'_2 &= b_2\exp (i\phi_\text{ARM}^b/2),\\
b'_0 &= b_0\exp (i\phi_\text{RC}^b/2) \label{app:b0__sol}\\
b_3 &\approx b'_0 r_1 + ib'_2k_{2020}t_1 \tag{\ref{eq:be_approx} Repeated}\\
b'_3 &= b_3 \exp (i\phi_\text{RC}^b/2).
\end{align}
There is no incoming $b_i$ field, so
\begin{align}
b_0 = r_0b'_3.
\end{align}
Solving this system of equations gives,
\begin{align}
    b_{0}=- \frac{i b_{1} k_{2020} r_{0} r_{2} t_{1} e^{\frac{i \left(\phi_\text{RC}^b + 2 \phi_\text{ARM}^b\right)}{2}}}{r_{0} r_{1} e^{i \phi_\text{RC}^b} - 1}. \label{app:b0_sol}
\end{align}
One may then substitute Eq.~\ref{app:b0_sol} into Eq.~\ref{app:b0__sol}, to get $b'_0$. Then one may substitute the result into Eq.~\ref{eq:b1} and solve to get Eq.~\ref{eq:b1sol}.
\section{General Comments on Modelling}
\label{app:general_comments_moddelling}
The numerical models in this work make use of \Finesse. A specific development version of \Finesse was used, identified as \texttt{3.0a2.dev7+g0ef5c64c.d20220113}. For modelling simplicity, \Finesse re-scales some phases so that cavities are resonant by default. Specifically, all distances are automatically rounded to an integer multiple of the wavelength. \Finesse can apply also arbitrary phase transforms to $k_{nmn'm'}$ which we disable as it can violate energy conservation (\S 4.10.2 \cite{finesse20}). The parameters can be set with
\begin{lstlisting}[language=Python]
import finesse
model = finesse.Model()
pc = model._settings.phase_config
pc.zero_k00 = False
pc.zero_tem00_gouy = True
\end{lstlisting}
The simple cavity simulation use $\lambda=$1.064\,\textmu m. The coupled-cavity simulations use $\lambda=$\,1.550\textmu m
\section{Coupled Cavity Model}
\label{app:cc_design}
The coupled cavity modelling in this work uses the Einstein Telescope Low Frequency nominal design~\cite{Rowlinson21}. The first mirror has a power transmission of 4.6\,\%, 37\,ppm of loss and 3505.6\,m RoC. We model a \textit{beamsplitter} placed 9\,m after the first mirror and having a power transmission of 0 and a loss of 37.5\,ppm. Thus the Y arm is disabled and not modelled. A beam expanding telescope is modelled in the arm, following the design of Rowlinson et. al. \cite{Rowlinson21}. This consists of a mirror with -50\,m RoC, 70\,m after the beamsplitter and a mirror with -82.5\,m RoC located 50\,m further. Both these optics have 287.5\,ppm loss. The ITM is a 20\,cm thick silicon mirror with a 75\,m focal length lens and 20\,ppm losses in the AR surface. The HR surface has transmission 7000\,ppm and 37.5\,ppm loss. The RoC is -5580. The ETM has 6\,ppm transmission, 37.5\,ppm loss and 5580 RoC. The cavity lenght is 10\,km and the wavelength is 1550\,nm. For further details see \cite{Rowlinson21,design_study_update_et}.
%
%
%
%
%
\section{Validation of $a_0$}
\label{app:discussion_a0}
\begin{figure}
    \centering
    \includegraphics[width=0.7\linewidth]{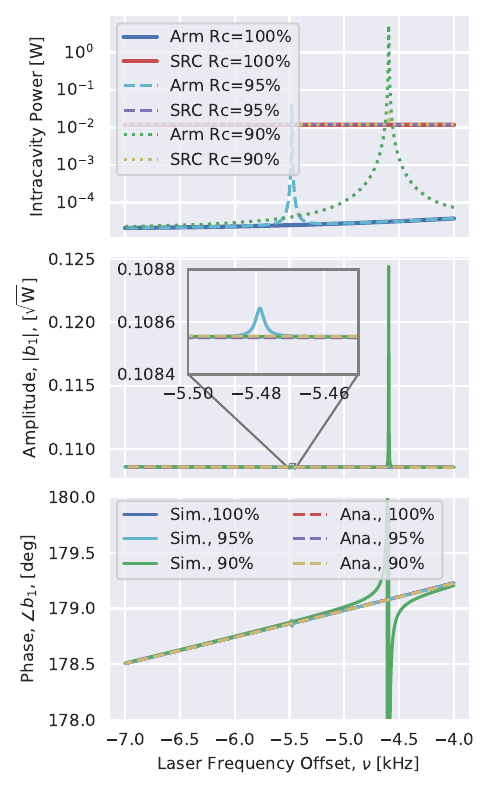} 
    \caption{The top plot shows the power measured in a \Finesse simulation. One can clearly see the resonances of the $b_1$ mode when the cavities are mismatched. The lower two plots show the amplitude and phase of the $a_0$ mode. The lower two plots share a legend. Solid lines are simulated results while dotted lines are computed with Eq.~\ref{eq:a0}. For a 5\,\% RoC mismatch, the inset shows a $\sim 0.1$\,\% deviation in amplitude and $\ll$0.1 deg. deviation in phase between the simulation and Eq.~\ref{eq:a0}. For a 10\,\% RoC mismatch, there is a significant phase change and 13\,\% deviation in amplitude. The deviations occurs when the approximations in \S~\ref{sec:cc} breakdown. See Appendix~\ref{app:discussion_a0} for more details. }
    \label{fig:a0}
\end{figure}
The approximations for $a_0$ neglect scatter from the HG20 mode into the HG00 mode. This is valid only when $k_{0000} a'_2 \gg k_{2000}b'_2$ and $k_{0000} a'_0 \gg k_{2000}b'_0$. The first equation is stating that the coupling from the HG00 mode in the arm to the HG00 mode in the RC, should be much stronger than any coupling from the HG20. Since the HG00 is not resonant in the arm (at the modulation frequency), but the HG20 is (at the modulation frequency), this approximation is not universally true. However, for $k_{2000}k_{0020} \ll 1$, only a small amount of power is being exchanged between the HG20 and HG00 modes. Thus, from power conservation, we know that the power returning from the resonance must be much less than the power driving the resonance.

The second equation is stating that coupling from the HG00 mode in the RC, into the arm must be much stronger than any coupling from the HG20 mode in the recycling cavity (at the modulation frequency). The HG20 is not resonant in the RC, whereas we assume the HG00 is within the FWHM of the RC. Therefore, the second constraint is usually satisfied when the $\fmodesep \lesssim \text{FWHM}_\text{RC}$. 

\Finesse defines power as $P = \sum_{\{\omega,n,m\}} |a_{\omega_i,n,m}|^2$ where $a_{\omega,n,m}$ is the field strength of a mode with frequency $\omega$ and spatial mode $u_{n,m}$. This power is shown in the top row of Fig~\ref{fig:a0} for the same set of simulation data as Fig~\ref{fig:b0}. Note that only the lower sideband is modelled here, the carrier field and PDH sidebands are omitted for simplicity. When there is no mismatch ($R_C=$100\,\%), $k_{0020}\rightarrow 0$ and so there is no power in the HG20 mode. As the mismatch increases, more power is present in the $b_1$ mode, as predicted by Eq.~\ref{eq:b1sol}. The lower two plots show the amplitude and phase of the mode $a_0$. Since the analytic equations are not modelling the \HG{20} mode scattering back into the \HG{00} mode, we do not expect them to change around the resonance. However, in reality when $k_{0020}$ increases our approximations break down, which is shown by the deviations between the analytic and numeric results for $5\,\%$ and $10\,\%$ mismatch.  10\,\% RoC mismatch causes the amplitude of the $a_0$ mode to be 13\,\% larger than Eq.~\ref{eq:a0} predicts. In addition, the phase of the HG00 mode in the RC is also modulated by the HG20 passing through resonance in the arm. A 5\,\% RoC mismatch also introduces a small deviation in both phase and amplitude of the $a_0$ mode.

\bibliography{main,bham-ifolab}

\end{document}